\documentclass[a4paper,11pt]{article}
\usepackage{jheppub}

\usepackage{graphicx}% Include figure files
\usepackage{bm}% bold math
\usepackage{hyperref}% 
\usepackage{amssymb}
\usepackage{amsmath}
\usepackage{amstext}
\usepackage{amsfonts}
\usepackage{amssymb}
\usepackage{subfigure}
\usepackage{color}
\usepackage{epsfig}
\usepackage{bbm}
\newcommand{\fm}{\mbox{fm}}

\newcommand{\Mev}{\mbox{MeV}}
\newcommand{\Gev}{\mbox{GeV}}

 \newcommand{\conf}{\mbox{conf}}
\newcommand{\plaq}{\mbox{pl}}
\newcommand{\rt}{\mbox{rt}}
\newcommand{\Tr}{\mbox{Tr}}
\newcommand{\sign}{\mbox{sign}}

\newcommand{\imp}{\mbox{imp}}
\newcommand{\ov}{\mbox{ov}}
\title{Tensor polarizability of the vector   mesons from $SU(3)$  lattice gauge theory}

 \author[a,b]{E.V. Luschevskaya}
 \author[c,a]{O.V. Teryaev}
           \author[a]{D.Yu. Golubkov}
    \author[a]{O.V. Solovjeva}
    \author[a,b]{R.A.Ishkuvatov}
 
\affiliation[a]{Institute for Theoretical and Experimental Physics named by A.I.Alikhanov of NRC ``Kurchatov Institute'', 117218,  Bolshaya Cheremushkinskaya  25, Moscow, Russia}
\affiliation[b]{Moscow Institute of Physics and Technology, Dolgoprudnyj, Institutskij lane 9, Moscow Region 141700, Russia}
\affiliation[c]{Joint Institute for Nuclear Research, Dubna, 141980, Russia} 

\emailAdd{luschevskaya@itep.ru}
\emailAdd{olga.solovjeva@itep.ru}
\emailAdd{teryaev@theor.jinr.ru}
\abstract{The magnetic dipole polarizabilities of the  vector $\rho^0$ and $\rho^{\pm}$  mesons in $SU(3)$ pure gauge theory are calculated in the article.  Based on this the authors explore the contribution of the dipole magnetic polarizabilities to the tensor polarization of the   vector   mesons in external abelian magnetic field. The tensor polarization leads to the dilepton asymmetry   observed in non-central heavy ion collisions and can be also estimated in lattice gauge theory.  
}

\keywords{
Strong magnetic field, quantum chromodynamics, lattice gauge theory, spin, magnetic dipole polarizability
}
\begin{document} 
\maketitle
\flushbottom

% \linenumbers

\section{Introduction}
\label{intro}

The influence of strong magnetic fields on quark-hadron matter represents  rich and full of surprising effects area of science.
These fields could exist in the Early Universe, they can influence the physics of neutron stars, and lead to non-trivial effects in non-central heavy-ion collisions in terrestrial laboratories.
 Behaviour of the hadron energy  in  external  magnetic field may provide information about the particle internal structure. In strong magnetic field the hadronic wave function deforms. This deformation  is defined  both by QED and QCD interactions inside the hadron. The magnetic polarizability and hyperpolarizabilities are quantities   describing the  response of the hadron to the external magnetic field, which we have explored in our previous work \cite{2017}. 
 Apparently, this non-linear response arises solely due to the strong QCD interaction binding quarks together.
 
  The magnetic field effect on the hadronic energy and structure was explored by theoretical models \cite{Simonov:2013,Simonov,Cho:2015,Taya:2015,Kawaguchi:2016,Hattori:2016,Gubler:2015} as well as lattice calculations \cite{Martinelli:1982, Liu:2015, Luschevskaya:2015a, Savage:2015, Luschevskaya:2015b, Bali:2015, Luschevskaya:2016}. The   polarizabilities of hadrons were also investigated by analytical methods  using dispersion relations in \cite{baldin,Filkov:2006}, the  magnetic moments were  studied in   \cite{Samsonov,Aliev,Djukanovic,Lee,Owen}.
  
  Below we discuss the calculations in  lattice gauge theory  with chiral invariant Dirac operator without dynamical quarks.  Our method enables to calculate the hadronic energies for different  spin projections on the  magnetic field axis. Magnetic polarizabilities and moments can be extracted as the fit parameters from the magnetic field value dependence of the energy.
The values of the magnetic polarizabilities depend on the meson spin projection   on the field axis. The physical meaning of this phenomenon is related to different deformations of a hadron  in the various space direction.  Therefore, the presence of a magnetic field creates a kind of anisotropy in space, which can lead to tensor polarization (alignment) of the vector meson and, after its decay, to dileptonic asymmetry in collisions of heavy ions.  
Dilepton anisotropy \cite{Teryaev} is an important physical characteristic which is sensitive to different   channels of particle's decay and can be utilized to disentangle contributions  of some channels.
Dilepton asymmetries provide information about the evolution of quark-gluon plasma in non-central heavy-ion collisions \cite{Baym:2017gzx}.   Here we study the influence of magnetic field on the tensor polarization of $\rho$ mesons which was detected in angular distribution of their decay products. This requires a more accurate calculation of polarizability, so, we repeat we repeat, improve and 
 extend   our previous analysis \cite{2017,Luschevskaya:2015a}.

\section{Technical details of the simulations}
 \label{sec-1}

We used ensembles of statistically independent $SU(3)$ gauge field configurations. For the generation of these configurations  the improved L\"uscher-Weisz action \cite{Luscher:1985} was used:
\begin{equation}
S=\beta_{\imp} \sum_{\plaq} S_{\plaq}-\frac{\beta_{\imp}}{20 u^2_0}\sum_{\rt}S_{\rt},
\end{equation}
where $S_{\plaq,\rt}=(1/3)\Tr(1-U_{\plaq,\rt})$ is the plaquette and
rectangular loop terms respectively, $u_0 = (W_ {1 \times 1})^{1/4} $ is defined by the relation
 $ W_{1 \times 1} = \langle (1/2) {\rm Tr} U_{\plaq} \rangle$ calculated at zero temperature \cite{Bornyakov:2005}.
 
  The calculations were carried out in $SU(3)$ lattice gauge theory without dynamical quarks.
We consider the lattice volumes $N_t \times N_s^3 = 18^4$ and $20^4$ and a set of   lattice spacings
$a=\{0.095, 0.105, 0.115\}\, \fm$. In   Table \ref{tbl:sim_det} we show the lattice volume $N_t\times N_s^3$, the lattice  spacing $a$, the corresponding $\beta_{\imp}$ values and the number of configurations.

%  \begin{table}[htb]
% \begin{center}
%\begin{tabular}{c|r|r|r|r}
%\hline
%$N_t \times N_s^3 $ & $ \beta_{imp}$  &$a \ fm$ & $ N_{conf} $  \\
%\hline
%$16^4$          &      8.20       & 0.115    & 245-250       \\
%\hline
%$18^4$          &      8.10       & 0.125    & 285-300         \\

%$18^4$          &      8.20       & 0.115    & 195-200         \\
%$18^4$          &      8.30       & 0.105    & 235-240          \\
%$18^4$          &      8.45       & 0.095    & 190-200          \\
%$18^4$          &      8.60       & 0.084    & 150-200          \\
%\hline
%$20^4$          &      8.20       & 0.115    & 270-300        \\
%$20^4$          &      8.45       & 0.095    & 190-200          \\
% \hline
%\end{tabular}
%\center{Table ~ 1. Parameters of ensembles of configurations of gluonic %fields.}
%\label{tbl:sim_det}
%\end{center}
%\end{table}

  \begin{table}[htb]
 \begin{center}
\begin{tabular}{|c|r|r|r|}
\hline
$N_t \times N_s^3 $ & $ \beta_{\imp}$  & $a, \ \fm$ & $ N_{\conf} $  \\
\hline
$18^4$          &      8.20       & 0.115    & 300         \\
$18^4$          &      8.30       & 0.105    & 400          \\
$18^4$          &      8.45       & 0.095    & 200          \\
$20^4$          &      8.20       & 0.115    & 300        \\
 \hline
\end{tabular}
\center{Table 1. The parameters and number of the lattice configurations used for the simulations.}
\label{tbl:sim_det}
\end{center}
\end{table}
In order to construct meson correlation  function,  we used quark propagators, which can be extracted as a series of  eigenfunctions and eigenvalues of discrete lattice version of the Dirac operator. 

\begin{equation}
D^{-1}(x,y)=\sum_{k<M}\frac{\psi_k(x) \psi^{\dagger}_k(y)}{i \lambda_k+m},
\label{lattice:propagator}
\end{equation}
where $ M = $ 50 is used for the calculations.

The next step is the numerical solution of the  Dirac equation and finding the eigenfunctions $\psi_k$ and the eigenvalues $\lambda_k$ of the Dirac operator for a quark located in an external gauge field 
$A_{\mu}$.
\begin{equation}
D \psi_k=i \lambda_k \psi_k, \  \ D=\gamma^{\mu} (\partial_{\mu}-iA_{\mu}) 
\label{Dirac}
\end{equation}
In these calculations we used the Neuberger overlap operator   \cite{Neuberger:1997}. This operator enables to consider the limit of massless quarks without breaking the chiral symmetry and can be written in the following form
 \begin{equation}
D_{\ov}=\frac{\rho}{a}\left( 1+D_W/\sqrt{D^{\dagger}_W D_W}  \right),
\label{overlap}
\end{equation}
where $D_W = M-\rho/a $ is the Wilson-Dirac operator with the negative mass parameter $\rho/a$, $a$ is the lattice spacing, $M$ is the Wilson term. 
The sign function $\sign(H_W)$ of the  Hermitian Wilson-Dirac operator $H_W$ is determined by the following relation
\begin{equation}
  \gamma_5 \sign(H_W)= D_W/ \sqrt{(D_W)^{\dagger} D_W}.
\label{sign_function}
\end{equation}
It is calculated using the min-max method of approximation by polynomials.

The fermionic fields satisfy the periodic boundary conditions in space and the antiperiodic boundary conditions in time.
We investigate the behaviour of the ground state energy of the meson in a gauge field, which is the sum of the  gluonic field and the external constant magnetic field. The magnetic field interacts only with quarks, so the magnetic field was added only into the Dirac operator.
A constant magnetic field $B$ is directed along the z-axis.
\begin{equation}
A_{\mu \, ij}\rightarrow A_{\mu \, ij} + A_{\mu}^{B} \delta_{ij},
\label{exchange}
\end{equation}
where
\begin{equation}
 A^B_{\mu}(x)=\frac{B}{2} (x_1 \delta_{\mu,2}-x_2\delta_{\mu,1}).
\end{equation}

The quantization of the magnetic field on a torus was originally discussed   in \cite{Hooft:1979,Zainuddin:1989,Chen:1996}. 
 In order to satisfy the boundary conditions for fermions, the magnetic field   has to be quantized  on the lattice   \cite{Al-Hashimi:2009}
 \begin{equation}
qB = \frac{2 \pi k} {(aN_s) ^ 2}, \ \ k \in \mathbb {Z},
\label{quantization}
\end{equation}
where $ q = - 1/3 \, e $, $aN_s$ is the lattice space extension.
 The calculation of the eigenfunctions and eigenvalues of the Dirac operator allows us to find the value of the two-point correlation function by means of which we determine the energy of the ground state.

 \section{Meson correlation functions}
 
We calculate the following correlation functions
\begin{equation}
\langle\psi^{\dagger}(x) \gamma_i \psi(x) \psi^{\dagger}(y) \gamma_j \psi(y)\rangle_A,
\label{observables}
\end{equation}
where $ \gamma_i, \gamma_j $ are  the Dirac gamma matrices with the    Lorentz indexes $ i, j = 1,2, 3 $ 
$ x = (\textbf {n} a, n_ta) $ and $ y = (\textbf{n}^{\prime} a, n^{\prime}_t a) $ are the coordinates on the lattice.

The spatial coordinates on the lattice  are written as follows:  $ \textbf{n}, \textbf{n}^{\prime} \in \Lambda_3 = \{(n_1, n_2, n_3) | n_i = 0,1, ..., N -1 \}$, $n_t, n^{\prime}_t$ are the numbers of sites in the time direction.
In the Euclidean space $\psi^{\dagger} = \bar{\psi}$.
The correlators of the $\rho^{\pm}$ mesons \eqref{observables} are computed with the use of the relation 
 \begin{equation}
\langle \bar{\psi}_{d,u}(x) \gamma_i \psi_{u,d}(x) \bar{\psi}_{u,d}(y) \gamma_j \psi_{d,u}(y) \rangle_A=-\Tr[\gamma_iD^{-1}_{u,d}(x,y)\gamma_jD^{-1}_{d,u}(y,x)],
\label{lattice:correlator}
\end{equation}
where   $D^{-1}_d$ and $D^{-1}_u$ are  the   propagators of the $d$ and $u$ quarks in the coordinate space. We consider the isospin symmetry case, so at zero magnetic field $D^{-1}_d=D^{-1}_u$, but at non-zero field this  
equality  is not fulfilled.

Since the $u$ and $d$ quarks interact differently with the field, the correlator    of the  $\rho^0$ meson is represented by the sum
\begin{equation}
\langle \bar{\psi}_{d}(x) \gamma_i \psi_{d}(x) \bar{\psi}_{d}(y) \gamma_j \psi_{d}(y) + \bar{\psi}_{u} (x)\gamma_i \psi_{u}(x) \bar{\psi}_{u}(y)\gamma_j \psi_{u}(y) \rangle_A=
\label{lattice:correlator_rho0}
\end{equation}
$$
-\Tr[\gamma_iD^{-1}_{d}(x,y)\gamma_jD^{-1}_{d}(y,x)]-\Tr[\gamma_iD^{-1}_{u}(x,y)\gamma_jD^{-1}_{u}(y,x)].
$$

We numerically perform   Fourier transform of the correlators  from the   coordinate space to the momentum space.
In order to find the ground-state energy, we consider mesons with zero spatial momentum.

The   ground-state energy of a meson with a definite spin projection  on the magnetic field axis is defined by the covariant density matrix.
We express the spin density matrix in terms of transverse ($e_x=(0,1,0,0), e_y=(0,0,1,0)$) and longitudinal ($e_z=(0,0,0,1)$) polarization vectors. Then the energies   of the $\rho^{\pm}$ and $\rho^0$ mesons with   the spin projection $s_z=0$  are obtained from equations \eqref{lattice:correlator} and \eqref{lattice:correlator_rho0} respectively after Fourier transform, where $i,j=3$.

 The combinations of the correlators
    \begin{equation}
C(s_z=\pm 1)=\langle O_1 (t)\bar{O}_1 (0)\rangle_A+\langle O_2 (t)\bar{O}_2 (0)\rangle_A \pm i(\langle O_1 (t)\bar{O}_2 (0)\rangle_A-\langle O_2 (t)\bar{O}_1 (0)\rangle_A)
\label{eq:CVV1}
    \end{equation}
    give the energies of vector mesons  with the   spin projections  equal to $+1$ and $-1$ on the field axis, where
 $O_1=\psi^{\dagger}_{d,u}(x) \gamma_1 \psi_{u,d}(x),\,  O_2=\psi_{d,u}^{\dagger}(x) \gamma_2 \psi_{u,d}(x)$ 
 are the interpolation operators of the $\rho^{\pm}$ mesons. 
 The interpolation operators for the $\rho^0$ case are constructed similarly to \eqref{eq:CVV1} taking into account \eqref{lattice:correlator_rho0}. 
 
The correlation function  can be expanded  in a series over the eigenstates of the Hamiltonian $\widehat{H}$\\
\begin{equation}
\langle O_i(t)  \bar{O}_j(0)\rangle_T=\frac{1}{Z}\sum_{m,n}\langle m|e^{-(T-t)\widehat{H}} \widehat{O}_i|n \rangle \langle n|e^{-t\widehat{H}} \widehat{O}_j^{\dagger}|m \rangle=
\label{series}
\end{equation}
$$= \frac{1}{Z}\sum_{m,n}e^{-(T-t)E_m}\langle m|\widehat{O}_i|n \rangle e^{-tE_n}\langle n| \widehat{O}_j^{\dagger}|m \rangle ,
$$
where $i,j=1,2,3$, $E_m$ and $E_n$ are the energies of the excited states with the numbers $m$ and $n$, and $Z=\sum_n \langle n|e^{-T\widehat{H}} |n \rangle=\sum_n e^{-TE_n}$  is the partition function. 

In   expression  \eqref{series}  we take out the factor $e^{-TE_0}$, as a result, we obtain the following relation\\
\begin{equation}
\langle O_i(t)  \bar{O}_j(0)\rangle_T=\frac{\sum_{m,n}e^{-(T-t)\Delta E_m}\langle m|\widehat{O}_i|n \rangle e^{-t \Delta E_n}\langle n| \widehat{O}_j^{\dagger}|m \rangle}{1+e^{-T \Delta E_1}+e^{-T \Delta E_2}+...},
\label{series2}
\end{equation}
where $\Delta E_n=E_n-E_0$.
In the thermodynamical  limit of the theory  $T\rightarrow\infty$  from   \eqref{series2}  we obtain
\begin{equation}
\langle O_i(t)  \bar{O}_j(0)\rangle_{T\rightarrow \infty}= \sum_{n} \langle 0|\widehat{O}_i|n \rangle \langle n| \widehat{O}_j^{\dagger}|0 \rangle e^{-t E_n}.
\end{equation} 
In our case we got the following expression\\
 $$
C(n_t)=\langle \psi^{\dagger}(\textbf{0},n_t) \gamma_i \psi(\textbf{0},n_t) \psi^{\dagger}(\textbf{0},0) \gamma_j \psi(\textbf{0},0)\rangle_A =
 $$
\begin{equation}
\sum_k\langle 0|\widehat{O}_i|n \rangle \langle n|\widehat{O}^{\dagger}_{j}|0 \rangle e^{-n_t a E_n}.
\label{sum}
 \end{equation}
  One can see from \eqref{sum} that  at-large  $n_t$ the main contribution comes from the ground state.  Taking into account periodic boundary conditions on the lattice we obtain the final formula  
 $$
C_{fit}(n_t)=A_0 e^{-n_t a  E_0} + A_0 e^{-(N_T-n_t)  a E_0}=
$$
\begin{equation}
2A_0 e^{-N_T a E_0/2} \cosh ((\frac{N_T}{2}-n_t) a E_0),
 \label{coshfit}
\end{equation}
where $A_0$ is the constant, $E_0$ is the ground state energy.
We use this formula for fitting the correlation functions obtained from  the lattice propagators. From this fits the ground state energies can be obtained.

 \section{Calculation of $m_{eff}$ and its statistical errors}

The  $\rho$-meson effective masses $m_{eff}$ were extracted from $\chi^2$-fits to the  standard
asymptotic parametrization \eqref{coshfit},
with fit parameters  being $m_{eff} = aE_0$ and $A_0$. 
The effective mass can be also found from the following equation:  
\begin{equation}
 \frac{C(n_t)}{C(n_t+1)}=\frac{\cosh(m_{eff}(n_t-N_T/2))}{\cosh(m_{eff}(n_t+1-N_T/2))},
 \label{effmass}
\end{equation}
in which the boundary conditions on the lattice are taken into account.
  We have checked that the results for the energies coincide  with the results obtained from  \eqref{effmass}, taking into account the correlation matrix between the adjacent points of the plateau.

In order to control the stability of  resulting $m_{eff}$ values, i.e. to determine the $m_{eff}(n_t)$ plateau,
the fits were performed in four ranges of $n_t = \left[\frac{N_T}{2}-k, \frac{N_T}{2}+k\right], \,\, k=1..4$,
where the fit quality could be satisfactory, $\chi^2/d.o.f. \sim 1$.
 Despite that the fit quality was often acceptable already at $k \simeq 4$,
the values of $m_{eff}$ at $k = 4$ could still be systematically overestimated.

\begin{figure}[htb]
\begin{center}
  \includegraphics[width=7.5cm,angle=-90]{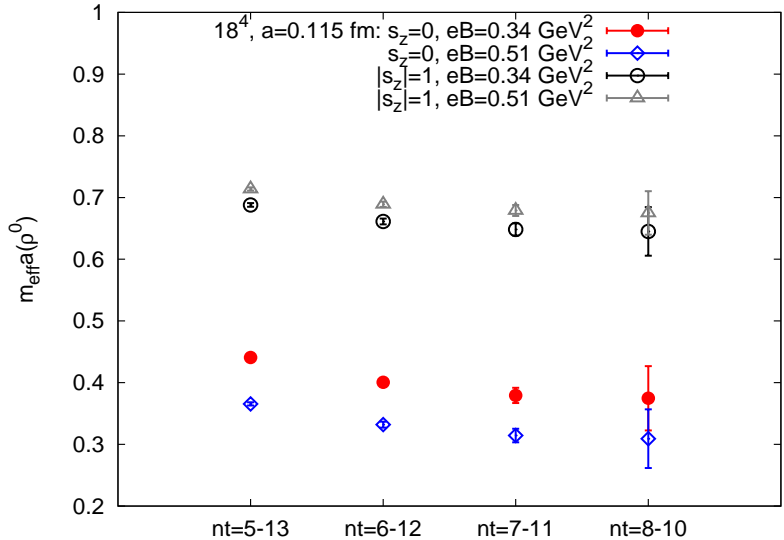}
  \caption{The $m_{eff}$ stability plot of the $\rho^0$-meson for the spin projections $s_z=0$ and $|s_z|=1$ on the magnetic field axis depending on the  $n_t$ fit ranges. The results were obtained on the lattice with the volume $18^4$, the lattice spacing $0.115\ \fm$ and the pion  mass $m_{\pi}=541\ \Mev$.}
  \label{Fig:m_eff_example}
\end{center}
\end{figure}

Usually, the $m_{eff}$  plateau was reached at $k=3$, i.e. $n_t=6\div12 \, (7\div13)$ for lattice volume $18^4 (20^4)$.
In some cases a narrower $n_t$ range was conservatively chosen after a visual inspection of the fits
to ensure that $m_{eff}$ was well within the plateau.
A typical graph of $m_{eff}(\rho^0)$   depending on the fit range $n_t$ is shown in Fig.~\ref{Fig:m_eff_example} for the lattice volume $18^4$, the lattice spacing $a=0.115\ \fm$, spin projections $|s_z|=0,1$ on the magnetic field axis  and two values of the magnetic field.

The uncertainties attributed to the resulting $\rho$-meson effective masses $m_{eff}$ are statistical-only,
as determined by the fit in the chosen $n_t$ range.

  \section{Magnetic  polarizabilities of the  $\rho^{\pm}$ mesons}
\label{sec-5}

 In the magnetic field the  energy levels of the point-like charged particle are described by the following dependency \cite{Kharzeevbook}
\begin{equation}
E^2=p_z^2+(2n+1)|qB|-gsqB+m^2, 
\label{eqLL}
\end{equation}
where $p_z$ is the momentum in the magnetic field direction,  $n$ is the
principal quantum number, $q$ is the particle electric charge, $g$ is the  g-factor, $s$ and $m$ are 
the  particle spin   and mass respectively.  

As we discussed earlier \cite{2017} the  magnetic field can affect the internal structure of vector mesons if it is sufficiently strong. This influence is characterized by non-zero magnetic polarizabilities and hyperpolarizabilities which depend  on the projection of the meson spin $s_z$ on the direction of the magnetic field. It means that the response of the fermionic currents inside the meson is defined by the mutual orientation of the quarks   and the external magnetic field.  These   phenomena are interesting and can give  a contribution to the polarization of the emitted charged particles   in the strong magnetic field which we discuss below.

According to the parity conservation for the  spin projection $s_z=0$  the energy squared gets corrections from the non-linear  terms  of even powers in a magnetic field. For  the spin projections $s_z=+1$ and $s_z=-1$ the energy squared contains the terms of both even and odd powers in the field.  We found  the contributions of these terms depend on the interval of magnetic fields considered. So, for the lattices considered the correction of   the third power term to the lowest energy sub-level with $qs_z=+1$ is not larger than $20\%$ and compatible with errors at $eB\in [0,1.2]\, \Gev^2$.  It may be seen from Fig. 9 presented in our previous work \cite{2017}. 
 Similarly, a correction of the fourth power does not give a significant contribution to the square of the energy for the case $s_z=0$.
 
\begin{figure}[htb]
\begin{center}
\includegraphics[width=7.5cm,angle=-90]{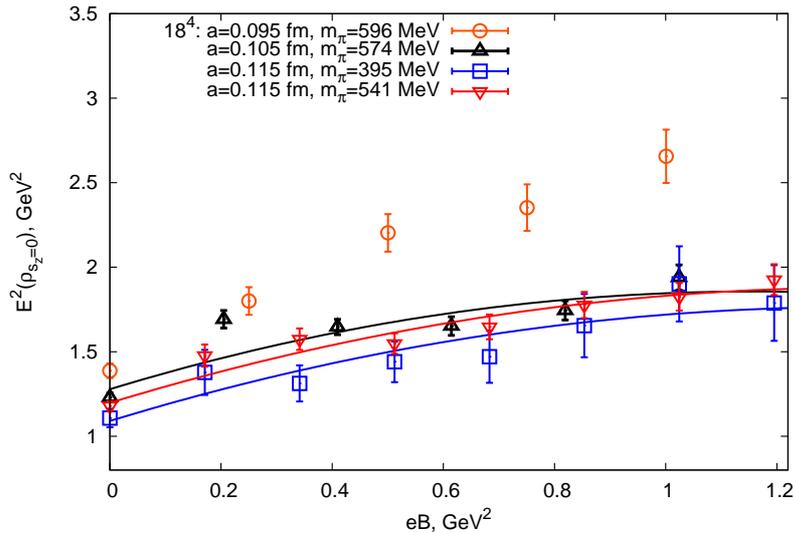}
\caption{The energy squared of the charged $\rho$ meson for the projection $s_z=0$  depending on the magnetic field value for various lattice sets of data.
The solid lines correspond to the  fits of  the lattice data obtained with the use of   formula   \eqref{eq:betas0}}
\label{Fig:rho:beta_s0}
\end{center}
\end{figure}

Therefore, we obtain the dipole magnetic polarizability for the spin projection $s_z=0$ from the fit of lattice data by the following relation
  \begin{equation}
E^2_{s_z=0} =|eB|+m^2 -4 \pi m \beta_m (eB)^2 
\label{eq:betas0}
\end{equation}
at  $eB\in[0,1.2]\ \Gev^2$, where $eB$ is the magnetic field in $\Gev^2$, $\beta_m$ is the dipole magnetic polarizability and fit parameter, $m$ is the fit parameter also. The  lattice results  together with the fitting curves are shown in Fig.~\ref{Fig:rho:beta_s0} which
 shows the energy squared increasing with the magnetic field value. At $eB\sim 0.2\div0.3\, \Gev^2$ we observe   a hump which leads to increase of the errors in determination of $\beta_m(s_z=0)$ value. 
  It can be a result of lattice spacing and lattice volume effects or have some physical explanation. 
 We cannot perform  a 	 rigorous analysis due to high errors of the calculations, but we note that this hump  is absent for the lattice with smaller lattice spacing $a=0.095\ \fm$. Therefore it is more probably a lattice spacing artefact.

The  values of the $\beta_m$   with the values of $\chi^2/n_{d.o.f.}$  and lattice parameters are shown in Table \ref{Table_s0}.

   \begin{table}[htb]
 \begin{center}
\begin{tabular}{c|r|r|r|r}
\hline
\hline
$V$             & $m_{\pi}(\Mev)$ &$a(\fm)$   & $\beta_m(\Gev^{-3})$    &   $\chi^2/d.o.f.$ \\
\hline
$18^4$           & $574 \pm 7$  & $0.105$   & $0.03\pm 0.01$        &  $6.90$ \\
\hline
$18^4$           & $395 \pm 6$  & $0.115$   & $0.028\pm 0.006$        &  $0.53$ \\
\hline
$18^4$           & $541 \pm 3$  & $0.115$   & $0.027\pm 0.004$        &  $1.25$ \\
\hline
\hline
\end{tabular}
\end{center}
\caption{The magnetic dipole polarizability $\beta_m$ of the charged $\rho$ meson with spin projection $s_z=0$ obtained  from the fits \eqref{eq:betas0} of the lattice data sets.  The pion masses  are represented in the second column, the lattice spacings in the third column, the values of $ \chi^2/d.o.f$   are shown in the last one.}
 \label{Table_s0}
 \end{table}
 
The  behaviour of the energy squared   for the   case  $qs_z=+1$ (which corresponds to $\rho^{-}$ at $s_z=-1$ and $\rho^{+}$ at $s_z=+1$)  can be described  by the dependency
  \begin{equation}
E^2_{qs_z=+1} =|eB| -g(eB) +m^2 -4 \pi m \beta_m (eB)^2 
\label{eq:betasp1}
\end{equation}
at  $eB\in[0,1.2]\ \Gev^2$.  The magnetic dipole polarizability was obtained from the fit of lattice data by the relation \eqref{eq:betasp1}, where $m$, $g$ and $\beta_m$ are the fit parameters.

\begin{figure}[htb]
\begin{center}
\includegraphics[width=7.5cm,angle=-90]{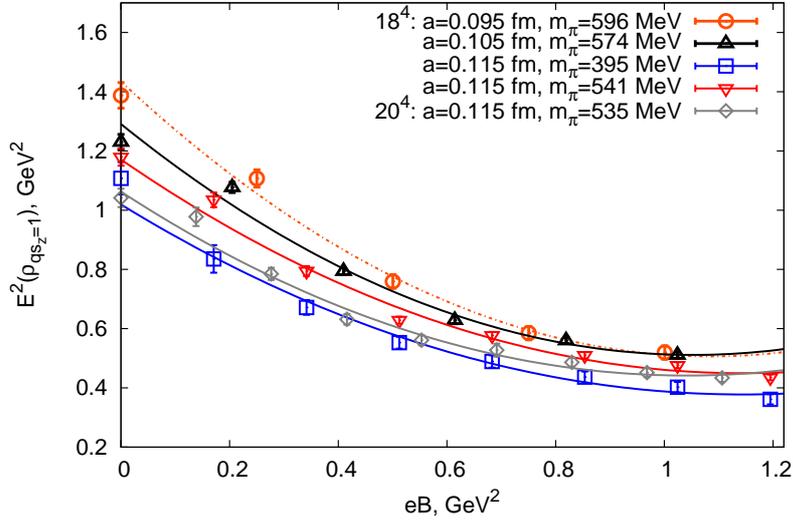}
\caption{The energy squared of the charged $\rho$ meson for the case $qs_z=+1$  depending on the magnetic field value for various lattice data sets. The solid lines correspond to the  fits of  the lattice data obtained with the use of   formula   \eqref{eq:betasp1}}
\label{Fig:rho:beta_s1}
\end{center}
\end{figure}

We represent this energy component in Fig.\ref{Fig:rho:beta_s1} for the lattice volume $18^4$, lattice spacings $0.105\ \fm$, $0.115\ \fm$  and for the lattice volume $20^4$, lattice spacing $0.115\ \fm$. The values of pion mass are also shown.  
 
The values of the magnetic dipole polarizability are represented in Table \ref{Table_s1}.    These results are obtained from the 3-parameter fit in comparison with the results of our previous paper (see Table 3 and Fig.9), where we fixed the $g$-factor value and used 2-parametric fit.  The  $\beta_m$ values are in a good agreement with our previous ones within the error range.

 \begin{table}[htb]
 \begin{center}
\begin{tabular}{c|r|r|r|r|r}
\hline
\hline
$V$              & $m_{\pi}(\Mev)$  &  $a(\fm)$ &  $g$-factor    & $\beta_m(\Gev^{-3})$      & $\chi^2/d.o.f.$\\
\hline
$18^4$           & $574 \pm 7$  &      $0.105$ &    $2.48 \pm 0.19$  & $-0.049\pm 0.010$     & $2.66$ \\
\hline
 $18^4$          & $541 \pm 3$   &      $0.115$ &   $2.26 \pm 0.14$  & $-0.041\pm 0.006$     & $2.32$ \\
\hline
 $20^4$          & $535 \pm 4$   &      $0.115$ &    $2.19\pm 0.12$   &  $-0.044\pm0.006$    & $1.48$ \\
\hline
$18^4$           & $395 \pm 6$   &      $0.115$ &    $2.12\pm 0.13 $   &  $ -0.039\pm 0.006$    & $1.49$  \\
\hline
\hline
\end{tabular}
\end{center}
\caption{The magnetic dipole moment and the magnetic dipole polarizability   of the charged $\rho$ meson with $qs_z=+1$ for  the lattice spacings $0.105\ \fm$, $0.115\ \fm$, the lattice volume    $18^4$, various pion masses and for the lattice spacing $0.115\ \fm$, the lattice volume $20^4$ and the pion mass $m_{\pi}=535(4)\  \Mev$  with their errors  and $ \chi^2/d.o.f$ values. The results were obtained with the use of 3-parametric fit \eqref{eq:betasp1} at $eB\in[0,1.2]$ shown in Fig.~\ref{Fig:rho:beta_s1}.}
\label{Table_s1}
 \end{table}

\begin{figure}[htb]
\begin{center}
\includegraphics[width=7.5cm,angle=-90]{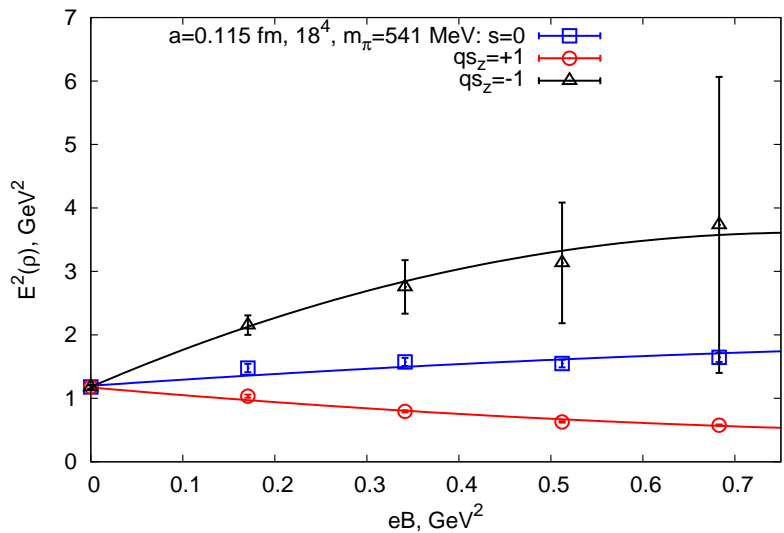}
\caption{The energy squared of the charged $\rho$ meson for the lattice volume $18^4$, lattice spacing $0.115\ \fm$ and the pion mass $m_{\pi}=541\ \Mev$ in dependence on the meson charge and spin projection on the magnetic  field axis.}
\label{Fig:rho:allspins}
\end{center}
\end{figure}

In Fig.\ref{Fig:rho:allspins} we represent the energy squared for different meson spin projections on the field axis for the lattice volume $18^4$, the lattice spacing $a=0.115\ \fm$ and the pion mass equal to $541\ \Mev$. 
The energy errors for the case $qs_z=-1$ are sufficiently high.  
For the case $qs_z=-1$ the fit was using the formula 
\begin{equation}
%E^2 =|eB| +g(eB) +m^2 -4 \pi m \beta_m (eB)^2 
E^2_{qs_z=-1} =|eB| +g(eB) +m^2 -4 \pi m \beta_m (eB)^2\,.
\label{eq:betasm1}
\end{equation}

 \begin{figure}[htb]
\begin{center}
\includegraphics[width=7.5cm,angle=-90]{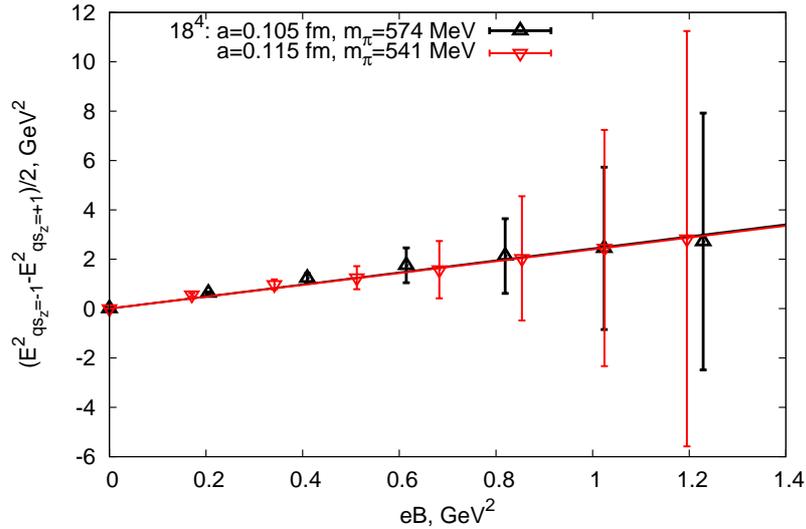}
\caption{The value of $(E^2_{qs_z=-1}-E^2_{qs_z=+1})/2$ versus the magnetic field value for the lattice volume $18^4$, lattice spacings $0.105\ \fm$ and $0.115\ \fm$.}
\label{Fig:rho:gfactor}
\end{center}
\end{figure}

To estimate the contribution of the magnetic dipole polarizability to the lepton asymmetry it is needed to know the dipole magnetic polarizability for the $s_z=+1$ and   $s_z=-1$ cases. 
However, due to big statistical errors   the magnetic dipole polarizability cannot be obtained from the energy dependence at $s_z=+1$ if we consider $\rho^-$ or from $s_z=-1$ if we consider $\rho^+$. 
Nevertheless, the parity conservation demands the equality of the magnetic dipole polarizabilities for $s_z=+1$ and $s_z=-1$. Then it follows from \eqref{eq:betasp1} and \eqref{eq:betasm1} 
\begin{equation}
E^2_{qs_z=-1}-E^2_{qs_z=+1}=2g(eB)\,.
\label{eq:diff}
\end{equation}

In order to make sure that the relation \eqref{eq:diff} is  satisfied for our data we represent  the value  of $(E^2_{qs_z=-1}-E^2_{qs_z=+1})/2$ versus the magnetic field value  in Fig.~\ref{Fig:rho:gfactor}. 
This quantity increases linearly with the magnetic field. The values of the $g$-factor obtained from these fits  are equal to $2.4\pm0.1$ for the lattice spacing $a=0.105\ \fm$ and $2.40\pm0.04$ for the spacing $a=0.115\ \fm$. The errors of the $g$-factor determination are underestimated because the effective mass plateau is very noisy for the $qs_z=-1$. We have determined the $g$-factor from the lowest energy sub-level in our previous work \cite{2017}.

 \section{Magnetic polarizability of the $\rho^0$ meson}
\label{sec-6}

We have discussed the magnetic polarizabilities of the neutral $\rho$ mesons  previously in \cite{Luschevskaya:2015a}, where we considered the toy model with   one type of quarks. In an external magnetic field the u- and d-quarks couple differently to the magnetic field. It has to be taken into account when we calculate the physical observables.
We repeat the analysis represented in \cite{Luschevskaya:2015a} with the difference that we find the energies from the  correlation function \eqref{lattice:correlator_rho0} and  increase the  statistics. 

In the relativistic case the  energy   of the  neutral  $\rho$ meson with spin projection $s_z=0$ is described by the following dependency
\begin{equation}
E^2 =m^2 -4 \pi m \beta_m (eB)^2-4 \pi m \beta_m^{h2} (eB)^4-4 \pi m \beta_m^{h4} (eB)^6-4 \pi m \beta_m^{h6} (eB)^8-...\,,
\label{eq:beta_rho0_s0}
\end{equation}
where $\beta_m^{h2}$, $\beta_m^{h4}$ and $\beta_m^{h6}$ are the various  magnetic hyperpolarizabilities of higher orders, $m$ is the mass of the meson at zero field.

   \begin{figure}[htb]
\begin{center}
\includegraphics[width=7.5cm,angle=-90]{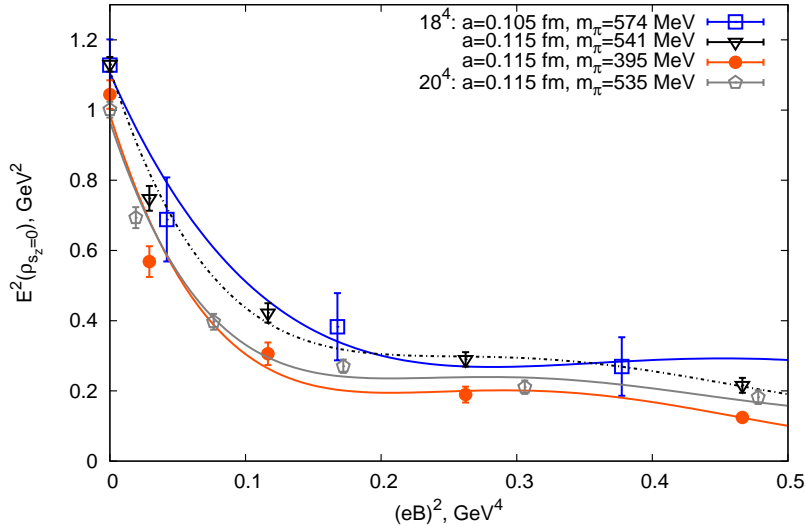}
\caption{The energy squared of the $\rho^0$ ($s_z$ = 0) ground state   versus the field value squared. The data are shown by points for various lattice spacings, pion masses and two lattice volumes $18^4$ and $20^4$ with the fits obtained using  formula \eqref{eq:beta_rho0_s0}.}
\label{Fig:rho:s0uudd}
\end{center}
\end{figure}

In Fig.~\ref{Fig:rho:s0uudd} we represent  the energy squared of the $\rho^0$ meson with $s_z=0$ versus the  field  squared for various lattices and pion masses at $(eB)^2\in[0:0.5]\, \Gev^4$. The lattice data are depicted by points.  We observe the energy decrease rapidly for all the lattice data. It also strongly depends   on the lattice spacing, the lattice volume, and the pion mass. 
The lines are the fits of these data obtained with the use of formula \eqref{eq:beta_rho0_s0}  when we also include terms $\sim (eB)^{10}$ and $\sim (eB)^{12}$. For the lattice with spacing $a=0.105\ \fm$ the fits were performed at $(eB)^2\in[0:1.7]\, \Gev^4$, for the other lattices we use  $(eB)^2\in[0:1.5]\, \Gev^4$.  
The terms of higher powers of field begin to contribute    significantly  at low fields. So, it is difficult to extract the magnetic polarizability,  because the quantization condition imposes a limitation on the minimal field value.
 We do not perform  an extrapolation to the chiral limit since  we are interested only  in the qualitative predictions at this stage of investigations.  

 In Table \ref{Table:rho0:beta_s0}   one can find the    magnetic polarizability $\beta_m$ and hyperpolarizability $\beta_m^{2h}$ obtained from the fits. The lattice volume $V$, the lattice spacing $a$,  the pion mass $m_{\pi}$, the interval of fields selected for the fitting procedure and $\chi^2/n.d.o.f.$ are also shown. The results agree with each other within the errors.

  \begin{table}[htb]
 \begin{center}
\begin{tabular}{c|r|r|r|r|r|r|r}
\hline
\hline
$V$ &   $a(\fm)$& $m_{\pi}(\Mev)$& $\beta_m(\Gev^{-3})$& $\beta_m^{2h}(\Gev^{-7})$ &$n$ & $(eB)^2(\Gev^4$)  &$\chi^2/d.o.f.$\\
\hline
$18^4$   &  $0.105$ & $574 \pm 7$         & $ 0.66\pm 0.16$  &  $ -2.51 \pm 0.98 $  & $10$ &  $[0:1.7]$     &$1.04  $ \\
\hline
 $18^4$  & $0.115$   & $541 \pm 3$       & $ 0.90\pm0.16$   &  $ -5.11 \pm 1.59 $    & $12$ &   $[0:1.5]$   &$ 2.46  $ \\
\hline
 $20^4$  & $0.115$  & $535 \pm 4$        & $ 0.95\pm0.15$    &  $  -5.78 \pm 1.60 $    & $12$ &  $[0:1.5]$     & $2.63  $ \\
\hline
$18^4$   & $0.115$  & $395 \pm 6$        & $ 0.98\pm 0.30 $  &  $ -5.79  \pm 2.74 $   & $12$ & $[0:1.5]$    & $ 3.32  $  \\
\hline
\hline
\end{tabular}
\end{center}
\caption{The value of the magnetic dipole polarizability $\beta_m$ and the magnetic hyperpolarizability  $\beta_m^{2h}$ for the $\rho^0$ with spin projection $s_z=0$ are shown for the various lattice volumes $V$, lattice spacings $a$ and pion masses $m_{\pi}$. The degree of polynomial $n$, field range used for the fitting and $\chi^2/d.o.f$ values are represented in columns sixth to eighth  correspondingly.}
\label{Table:rho0:beta_s0}
 \end{table}
 
 \begin{figure}[htb]
\begin{center}
\includegraphics[width=7.5cm,angle=-90]{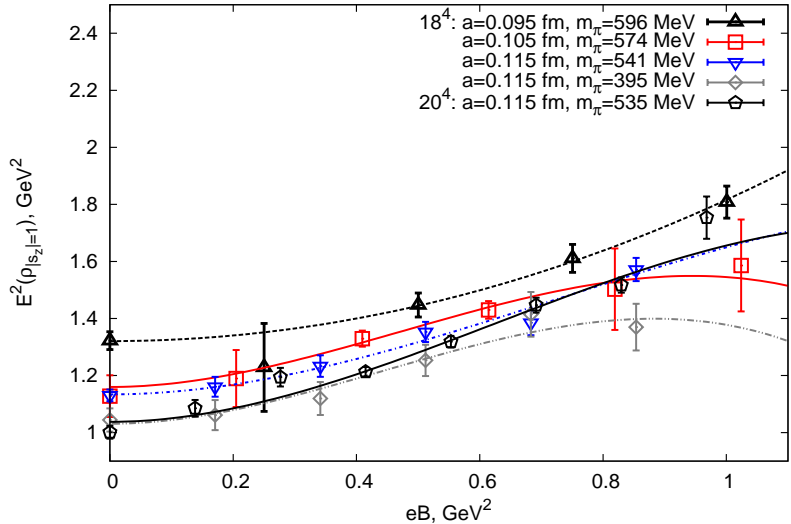}
\caption{The energy squared of the $\rho^0(|s_z|=1)$ meson versus magnetic field for various lattice spacings, pion masses, and two lattice volumes $18^4$ and $20^4$.    
The lines correspond to the fits of the lattice data obtained using formula \eqref{eq:beta_rho0_s1}.}
\label{Fig:rho:s1uudd}
\end{center}
\end{figure} 

At $eB\in[0:1.2]\ \Gev^2$ for the spin projection $|s_z|=1$ on the field axis the energy squared of the neutral vector meson can be described by the following relation:
\begin{equation}
E^2 =m^2 -4 \pi m \beta_m (eB)^2-4 \pi m \beta_m^{h1} (eB)^3\,.
\label{eq:beta_rho0_s1}
\end{equation}
In Fig.~\ref{Fig:rho:s1uudd} the energy squared is shown for the $\rho^0$ meson with the spin projection $|s_z|=1$, the energies of the neutral vector meson for the $s_z=+1$ and $s_z=-1$ coincide due to the conservation of C-parity.
The magnetic polarizability $\beta_m$ is obtained  from the fit of the lattice data by formula \eqref{eq:beta_rho0_s1} for the lattices with spacings $0.105\ \fm$ and $0.115\ \fm$, where $m$, $\beta_m$ and $\beta_m^{h1}$ are the fit parameters.
The lattice data for $a=0.084\ \fm$ and $a=0.095\ \fm$ do not allow to extract statistically significant $\beta_m$ values, but we show them to check the lattice volume and lattice spacing effects.
 The $\beta_m$ values with the errors and other parameters  are shown  in Table \ref{Table:rho0:beta_s1}. The results    agree with each other within the errors.

   \begin{table}[htb]
 \begin{center}
\begin{tabular}{c|r|r|r|r|r|r}
\hline
\hline
$V$    &  $a(\fm)$  & $m_{\pi}(\Mev)$   & $\beta_m(\Gev^{-3})$& $\beta_m^{1h}(\Gev^{-5})$ &     $eB,\ \Gev^2$    & $\chi^2/d.o.f.$\\
\hline
$18^4$  &  $0.105$ & $574 \pm 7$            & $ -0.10 \pm 0.02 $  &  $ 0.07 \pm 0.02$  &     $ [0:1.1]  $  & $0.47$ \\
\hline
 $18^4$ & $0.115$ & $541 \pm 3$            & $ -0.07 \pm 0.02 $  &     $ 0.03 \pm 0.03$ &        $ [0:1.1]  $  & $0.87$ \\
\hline
 $20^4$ & $0.115$ & $535 \pm 4$            & $ -0.10 \pm 0.02$    &      $0.06 \pm 0.03$  &     $ [0:1.1]  $  & $1.54$ \\ 
\hline
$18^4$ & $0.115$  & $395 \pm 6$           & $ -0.11\pm 0.03  $  &      $ 0.08 \pm 0.03$  &     $ [0:1.1]  $   & $0.65  $  \\
\hline
\hline
\end{tabular}
\end{center}
\caption{The value of the magnetic dipole polarizability $\beta_m$ and the magnetic hyperpolarizability  $\beta_m^{1h}$ are shown for $\rho^0$ with $|s_z|=1$, for  the  lattice volumes $18^4$ and $20^4$, the lattice spacings $0.105\ \fm$ and $0.115\ \fm$ and various pion masses. The fourth and the last columns contain the intervals of the magnetic field used for the fit and $\chi^2/d.o.f.$  correspondingly.}
\label{Table:rho0:beta_s1}
 \end{table}

 \section{Tensor magnetic polarizability}
 \label{sec-7}
 
 In the previous sections we have found that the energy of the $\rho^0$ meson with $s_z=\pm 1$   increases versus the magnetic field value, while the energy of the $\rho^0$ meson with $s_z=0$   diminishes   quickly versus the magnetic field value.
For the $\rho^{\pm}$ the energy decreases for the case $qs_z=+1$ and increases for the $qs_z=0$ and $qs_z=-1$ cases.

Such energy behaviour has some physical consequences.
The small energy is more profitable than the high energy, so the longitudinal polarization of the $\rho^0$ meson  corresponding to $s_z=0$ has to dominate in the collisions.
Therefore, in the non-central heavy-ion collision the magnetic field favours longitudinal polarization of the $\rho^0$ mesons.

The dilepton asymmetries in non-central heavy ion collisions depend on the energy behaviour of the vector mesons.
According to the vector dominance principle the vector mesons can directly convert to virtual photons.
In turn, the electromagnetic decay of virtual photons is one of the main sources of dilepton production in heavy-ion collisions.
When dileptons are produced in non-central heavy-ion collisions,
their anisotropy depends on the response of the spin structure of intermediate resonances, such as $\rho^0$
to the magnetic field of the collision.
Therefore, it is of particular interest to  identify and distinguish between various sources of dileptons emitted from non-central heavy-ion collisions.

The shape of the dilepton distribution is characterized by the following differential cross section:
\begin{equation}
 \frac{d\sigma}{dM^2 d\cos\theta}=A(M^2)(1+B\cos^2 \theta),
 \label{cross_sec}
\end{equation}
where $M^2=(p_1+p_2)^2$ is the energy of the lepton pair in their rest frame, $p_1$ and $p_2$ are the four-momenta of the leptons,  $\theta$ is the angle between the momenta of the virtual photon and  the lepton. 
 The asymmetry coefficient $B$ is defined by the polarization of the virtual photons produced in the collisions:
\begin{equation}
 B =\frac{\gamma_{\perp}-\gamma_{\parallel}}{\gamma_{\perp}+\gamma_{\parallel}},
\label{asymm}
\end{equation} 
where the  $\gamma_{\perp,\parallel}$ are the contributions of the transverse and
longitudinal polarizations of the virtual intermediate photon.

Our lattice calculations make possible to get the estimation of the asymmetry factor for such processes.  
For the vector particle in Cartesian basis the polarization  tensor has the following form
\begin{equation}
 P_{ij}=\frac{3}{2}\langle s_i s_j+s_j s_i \rangle -2 \delta_{ij}.
\end{equation}
If $w_{s_z=+1}$, $w_{s_z=-1} $ and $w_{s_z=0}$ are the probabilities that the $\rho$ meson has a spin  projection on the field direction equal to $+1$, $-1$ and $0$ correspondingly,
then  the value of the component $P_{33}$ can be represented in terms of these probabilities in the following form:
\begin{equation}
 P_{33}=w_{s_z=+1}+w_{s_z=-1}-2w_{s_z=0}=\frac{N_{s_z=+1}+N_{s_z=-1}-2N_{s_z=0}}{N_{s_z=+1}+N_{s_z=-1}+N_{s_z=0}},
\label{asymmetry_coef}
\end{equation} 
where $N_{s_z=+1}$, $N_{s_z=-1}$ and $N_{s_z=0}$ are the numbers of particles with different spin projections.

As $w_{s_z=+1}+w_{s_z=-1}+w_{s_z=0}=1$, therefore, $P_{33}=1-3w_{s_z=0}$ and $-2 \leq P_{33}\leq 1$.   
  The tensor polarizability describes the effect of the magnetic field for the spin states of the $\rho$ meson, in turn, the spin states and tensor polarization is revealed in its decay to lepton pair, see \cite{Teryaev}. In a strong magnetic field  the spin of the particle  tends to align along the field direction, but the non-zero temperature leads to the spin-flipping.

  The differential cross section  for the decay of $\rho$ meson to the lepton pair  may be written as
  \begin{equation}
 \frac{d\sigma}{dM^2 d\cos\theta}=N(M^2)(1+\frac{1}{4}P_{33}(3\cos^2 \theta-1)).
 \label{cross_sec_2}
\end{equation}

 Comparing with \eqref{cross_sec} one can see that 
\begin{equation}
B=\frac{3P_{33}}{4-P_{33}}.
\end{equation}
For the transversely polarized $\rho$ meson $B=1$ and for the longitudinally polarized $B=-1$.

We also introduce  the tensor polarizability 
 \begin{equation}
 \beta_t=\frac{\beta_{s_z=+1}+\beta_{s_z=-1}-2\beta_{s_z=0}}{\beta_{s_z=+1}+\beta_{s_z=-1}+\beta_{s_z=0}},
 \end{equation}
  which is the measure of the  magnetic field effect on a vector  meson, in particular for high magnetic fields and temperature  $P\sim \beta_t$.

 We calculate $\beta_{t}$ on the lattice taking into account the equality $\beta_{s_z=+1}=\beta_{s_z=-1}$. The results are presented in Table \ref{betarho0} for the neutral $\rho$-meson and in Table   \ref{betarho} for the charged 
 $\rho$-meson. 
 The large negative values of $\beta_t$ suggest the dominating longitudinal polarization of the $\rho^0$-meson. The dileptons are mainly emitted in the directions perpendicular to the magnetic field axis.
This is a convincing result, as we clearly see from Figure \ref{Fig:rho:s0uudd} and \ref{Fig:rho:s1uudd},that the energy of the state with $s_z=0$ decreases, and the energy of $|s_z|=1$ increases.

 It was found previously that for soft dileptons the longitudinal polarization also dominates \cite{Polikarpov}.  This result was obtained by comparing the formation of soft dileptons with a nonzero component of the conductivity of the strongly interacting matter, parallel to the external magnetic field  \cite{Luschevskaya}.

    \begin{table}[htb]
 \begin{center}
\begin{tabular}{c|r|r|r}
\hline
\hline
$V$       &    $a(\fm)$    & $m_{\pi}(\Mev)$  &         $\beta_t$       \\
\hline    
$18^4$    &   $0.105$    & $574 \pm 7$  &            $-3.3\pm 0.6 $      \\
\hline
 $18^4$    &   $0.115$   & $541 \pm 3$   &            $-2.6\pm 0.2$    \\
\hline
 $20^4$     &  $0.115$   & $535 \pm 4$   &            $-2.8 \pm 0.3 $      \\
\hline
$18^4$       & $0.115$   & $395 \pm 6$   &               $-2.9\pm 0.5$   \\
\hline
\hline
\end{tabular}
\end{center}
\caption{The tensor polarizability $\beta_t$ of the $\rho^{0}$ meson is shown in the last column for the lattice volume $V$, the lattice spacing $a$ and the pion mass $m_{\pi}$.}
\label{betarho0}
 \end{table}
 
  \begin{table}[htb]
 \begin{center}
\begin{tabular}{c|r|r|r}
\hline
\hline
%$V$              & $m_{\pi}(\Mev)$  &  $a(\fm)$ &       $\beta_{t}$       \\
%\hline
%$18^4$           & $574 \pm 7$  &      $0.105$ &       $2.3\pm 0.7$      \\
%\hline
% $18^4$          & $541 \pm 3$   &      $0.115$ &      $2.5\pm 0.5$    \\
%\hline
% $20^4$          & $535 \pm 4$   &      $0.115$ &      $ 5\pm 2$      \\
%\hline
%$18^4$           & $395 \pm 6$   &      $0.115$ &         $2.7\pm 0.7$   \\
%\hline
%\hline
 $V$             & $a(\fm)$	& $m_{\pi}(\Mev)$ &       $\beta_t$       \\
\hline
 $18^4$          & $0.105$ 	& $574 \pm 7$     &       $2.3\pm 0.7$      \\
\hline
 $18^4$          & $0.115$ 	& $541 \pm 3$     &       $2.5\pm 0.5$    \\
\hline
 $18^4$          & $0.115$ 	& $395 \pm 6$     &       $2.7\pm 0.7$   \\
\hline
\hline
\end{tabular}
\end{center}
\caption{The tensor polarizability  $\beta_t$ of the $\rho^{\pm}$ is represented for various lattices and pion masses.}
\label{betarho}
 \end{table}

\section{Conclusion}

In this paper the calculation of the magnetic dipole and tensor polarizabilities have been presented for the charged and neutral $\rho$ mesons on the lattice volume. We perform a thorough analysis of the behaviour of the effective mass plateau depending on the number of points used for the fitting by hyperbolic cosine function.

In Section \ref{sec-5}  the magnetic dipole polarizability and $g$-factor for the charged mesons have been represented. We include the additional sets of data into consideration and increase statistics  with the lattice spacing  $a=0.105\ \fm$ and volume $18^4$.    This allows one to obtain the more statistically significant values of the magnetic dipole polarizabilities for the spin projection $s_z=0$, see Table \ref{Table:rho0:beta_s0}. 
In contrast to the previous results \cite{2017}, we have calculated the dipole magnetic polarizability for the case $qs_z=1$ from the 3-parametric fit \eqref{eq:betasp1}, where  the $g$-factor is a  free parameter  of the fit.  Obviously, we could find the $g$-factor values at smaller field interval and use these values for the  $\beta_m$ determination, but the purpose of this work was to obtain the qualitative predictions.

In Section \ref{sec-6} the dipole magnetic polarizability have been extracted for the case of the neutral $\rho$ meson. The magnetic dipole polarizability was calculated, taking into account that the $u$ and $d$-quarks couple differently to the magnetic field. In \cite{Luschevskaya:2015a} the magnetic dipole polarizability of $\rho^0$ was calculated in the toy model with only one type of quark. Section \ref{sec-7} is devoted to the discussion of the polarization of the dileptons which result from the decays of $\rho$ mesons.
 We have found that the longitudinal polarization of the $\rho^0$ mesons dominates in the collisions, because the low energy is more preferential than high energy. Therefore the dileptons occurring due to decays of the $\rho^0$ mesons will be emitted perpendicular to the direction of the magnetic field. 
  This result reinforces the previous results obtained in \cite{Luschevskaya}, i.e. the nonzero conductivity of the quark-hadronic matter in a strong magnetic field. 
  We introduce   new characteristics of the meson magnetic properties - the tensor palarizability. This quantity has been suggested to be related to the coefficient of asymmetry in the differential cross section for the dilepton production.

\acknowledgments

The authors are grateful to FAIR-ITEP supercomputer center where these numerical calculations were performed.
 This work   is completely supported by the grant from the Russian Science Foundation (project number 16-12-10059).

\end{document}